\documentclass[a4paper]{article}

\title{Numerical Models of Blackbody-Dominated GRBs}

\date{}
\author{Carlos Cuesta-Mart\'inez\\ }
	
\usepackage{graphicx}
\usepackage{hyperref}
\usepackage{amsmath}
\usepackage{float}

\begin{document}

\begin{flushleft}
{\large \bf Numerical Models of Blackbody-Dominated GRBs}\\[0.3cm]

{\bf C. Cuesta-Mart\'inez \footnote{Speaker} }\\ 
\small{\textit{Departamento de Astronom\'ia y Astrof\'isica, Universidad de Valencia, E-46100 Burjassot (Valencia), Spain}}\\
\emph{E-mail:} \texttt{carlos.cuesta$@$uv.es}\\[0.3cm]
 
{\bf Miguel \'Angel Aloy, Petar Mimica}
\\[0.0cm]
\small{\textit{Departamento de Astronom\'ia y Astrof\'isica, Universidad de Valencia, E-46100 Burjassot (Valencia), Spain}}\\[0.3cm]

{\bf Christina Th\"one}
\\[0.0cm]
\small{\textit{Instituto de Astrof\'isica de Andaluc\'ia (IAA-CSIC), Glorieta de la Astronom\'ia s/n, E-18008 Granada, Spain}}\\[0.3cm]

{\bf Antonio de Ugarte-Postigo}
\\[0.0cm]
\small{\textit{Instituto de Astrof\'isica de Andaluc\'ia (IAA-CSIC), Glorieta de la Astronom\'ia s/n, E-18008 Granada, Spain\\	
        Dark Cosmology Centre, Niels Bohr Institute, Juliane Maries Vej 30, DK-2100 Copenhagen, Denmark}}\\[0.5cm]
\end{flushleft}

Blackbody-dominated (BBD) gamma-ray bursts (GRBs) are events characterized by the absence of a typical afterglow, long durations and the presence of a significant thermal component following the prompt gamma-ray emission. GRB 101225A (the `Christmas burst') is a prototype of this class. A plausible progenitor system for it, and for the BBD-GRBs, is the merger of a neutron star (NS) and a helium core of an evolved, massive star. Using relativistic hydrodynamic simulations we model the propagation of an ultrarelativistic jet through the enviroment created by such a merger and we compute the whole radiative signature, both thermal and non-thermal, of the jet dynamical evolution. We find that the thermal emission originates from the interaction between the jet and the hydrogen envelope ejected during the NS/He merger.

\begin{flushleft}
{Swift: 10 Years of Discovery,\\}
{2-5 December 2014\\}
{La Sapienza University, Rome, Italy }
\end{flushleft}

\section{Introduction}

Thanks to the \emph{Swift} satellite, gamma-ray bursts (GRBs) have been studied in great detail during the last decade. 
Nowadays it is well accepted that the majority of GRBs can be divided into two different groups based in their durations
(long and short) \cite{kouveliotou93}, and that each class may arise from a different progenitor system. The long ones
(LGRBs), those which last more than 2 s, are well understood thank to the multiple afterglow detections up to date. The
temporal and spectral evolution can be modeled employing power laws, which indicates that the radiation is due to
non-thermal processes, as synchrotron emission.  LGRBs are thought to form in collapsars: stellar-mass black holes
sorrounded by thick accretion disks able to power ultrarelativistic jets.  However a few outliers of the two standard
classes have been detected over the past few years, challenging the standard long/short division. GRB 101225A is one
of these outliers.

\subsection{The `Christmas burst'}
GRB 101225A was detected by \emph{Swift} on 25th December 2010, earning the nickname `Christmas burst' (CB) by which it
is also known \cite{T11}. The observed duration was initially claimed to be $\sim 2000$s, but the estimate was later
increased to $\sim 7000$ s \cite{L14} since the burst was active during more than one \emph{Swift} orbit. After several
different redshift estimations \cite{T11,campana11}, Levan et al. \cite{L14} determined the redshift to be $z = 0.847$,
unequivocally demonstrating its cosmological nature. One of the unusual features of the CB is the presence of a thermal
component in its optical and X-ray spectrum. The best fit for the optical evolution is the emission from an expanding,
cooling blackbody. The X-ray component is well fitted considering an absorbed power-law spectrum with a blackbody
component (a thermal hotspot with a characteristic temperature of 1 keV).  Because of its extreme long duration the CB
has been suggested as a member of the subclass of {\em ultralong} GRBs. On the other hand, the presence of a blackbody
spectral component has placed the CB as a prototype of another subclass of bursts, the blackbody-dominated GRBs
(BBD-GRBs).  Other GRBs, such as GRB 060218, have been found with similar durations and thermal components.

\subsection{The NS/He merger scenario: a viable progenitor for BBD-GRBs}
The existence of non-standard GRBs shows that, beyond the collapsar model, there may exist other evolutionary channels
and ways of producing very long bursts. For GRB 101225A and BBD-GRBs Th\"one et al.\,\cite{T11} proposed an alternative
scenario based on the merger of a neutron star (NS) and helium star \cite{fryerwoosley98,zhangfryer01}. In this scenario a NS spirals into its massive
companion which undergoes a common envelope (CE) phase due to tidal forces. During the CE phase the outer shells of the
massive star are expelled into the external medium (mostly in the equatorial plane), with roughly the escape velocity,
thus creating a high-density enviroment. This debris forms the so-called CE shell of the system \cite{PaperI}. Eventually the
NS will merge with the core of the companion and form a black hole/disk system (or even a magnetar),
able to power an ultrarelativistic jet which will interact with the surrounding CE shell.

\section{Numerical method}

We aim to test whether the NS/He merger is a viable scenario for producing such anomalous GRBs by means of numerical
simulations. We will not focus on the complete evolution of the system but on the hydrodynamical evolution of an
ultrarelativistic jet, formed after the merger, and its interaction with the circumburst medium. Afterwards we will
calculate the synthetic emission from the hydrodynamical models and compare with real observations of the
CB. 

In these proceedings, we describe the setup and results of our reference model (RM). Interested readers are encouraged
to review the complete references of our work \cite{PaperI,PaperII}, where we perform an exhaustive parametric study of
all the elements of the system.

\subsection{Hydrodynamical setup}
For the hydrodynamical evolution we use the finite volume, HRSC relativistic hydrodynamics (RHD) code MRGENESIS \cite{aloy99,mimica09}, in 2D spherical coordinates assuming that the system is axisymmetric. 

The numerical grid has a resolution $n_r \times n_\theta = 5400 \times 270$ cells and is filled with a uniform external
medium of density $\rho_{\rm ext} = 8 \times 10^{-14}$ g cm$^{-3}$ and pressure $p_{\rm ext} =10^{-5} \rho_{\rm ext}
c^2$. In the radial direction the grid starts at a distance $R_{\rm 0} = 3 \times 10^{13}$ cm and stops at $R_f = 3.27
\times 10^{15}$ cm. The angular coordinate spans the range [0$^\circ$, 90$^\circ$]. Reflective boundary conditions are
imposed at $R_0$, the symmetry axis of the system and at the equator.  At a distance of $R_{\rm CE,in} = 4.5 \times
10^{13}$ cm we place a uniform, high-density shell which extends up to $R_{\rm CE,out} = 1.05 \times 10^{14}$ cm. The
shell density is $\rho_{\rm sh,CE} = 1.2 \times 10^{-10}$ g cm$^{-3} = 1500 \rho_{\rm ext}$, equivalent to a mass of
$M_{\rm CE,sh} \sim 0.26 M_\odot$. The pressure is uniform and matches that of external medium, $p_{\rm CE,sh} = p_{\rm
  ext}$. This shape mimics a torus with a low-density funnel (made of external medium) around the symmetry axis which
extends from $\theta_{\rm CE,in} = 1^\circ$ at $r=R_{\rm CE,in}$ to $\theta_{\rm CE,in} = 30^\circ$ at $r=R_{\rm
  CE,out}$. Since $R_{\rm CE,in} > R_0$ a gap filled of external medium is formed between these two regions. This gap is
artificial and is created with the only purpose to let the jet accelerate before the interaction with the CE shell
occurs. Furthermore we have checked that the presence of this gap has a negligible impact in our results \cite{PaperI}.
At $R_{\rm 0}$ an ultrarelativistic jet with an initial Lorentz factor $\Gamma_{\rm i} = 80$ and a specific enthalpy
$h_{\rm i} = 5$ is injected for $t_{\rm inj} = 7000 / (1 + z)$ s. This is equivalent to an asymptotic Lorentz factor of
$W_{\infty} = 400$. The injection is done in two stages: (1) constant up to $2000 / (1 + z)$ s and (2) decaying with
$t^{-5/3}$ up to $t_{\rm inj}$. After that the injection is not switched off abruptly but goes with $t^{-4}$ for reasons
of numerical stability. The isotropic energy $E_{\rm iso} = 4 \times 10^{53}$ erg is consistent with the observed lower
bound of $E_{\rm iso, \gamma + X} > 1.2 \times 10^{52}$ erg. The half-oppening angle of the jet is $\theta_{\rm j} =
17^\circ$, fulfulling that $\theta_{\rm j} > \theta_{\rm CE,in}$ and ensuring the interaction between the jet and the CE
shell. This values give a total jet energy $E_{\rm j} = E_{\rm iso} (1 - \cos{\theta_{\rm j}}) / 2 = 8.7 \times 10^{51}$
erg.

\subsection{Computing the emission}
For computing the synthetic electromagnetic emission from the RHD models we use an improved version of the radiative
transport code \emph{SPEV} \cite{mimica09}. We consider two different kind of emission processes: (1) non-thermal
radiation coming from electrons accelerated at shocks by stochastic magnetic fields, i.e. synchrotron radiation, and (2)
thermal radiation via free-free bremsstrahlung. In the former, we need to specify a set of non-thermal microphysical
parameters. We chose an `effective' fraction of the internal energy for the electrons $\epsilon'_{\rm e} \approx
10^{-2}$, and for the stochastic magnetic field of $\epsilon_B = 10^{-6}$ (see \cite{PaperII} for further details). The
electron spectral index is set to $p=2.3$. In the latter process, we consider that the ratio between emissivity and
absorptivity gives the blackbody intensity.

\section{Dynamical evolution}

\begin{figure}[h]
\includegraphics[height=.25\textwidth]{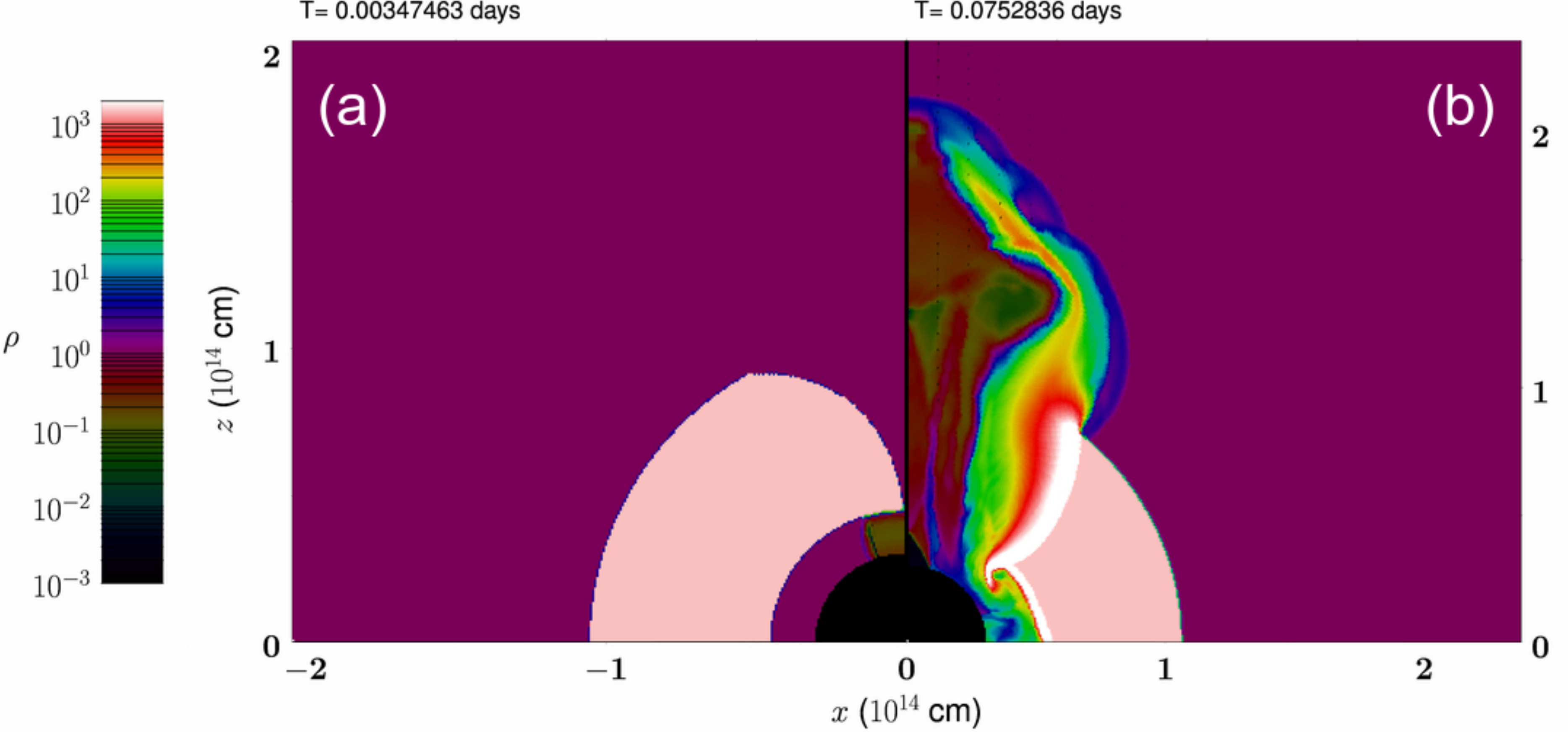}
\includegraphics[height=.25\textwidth]{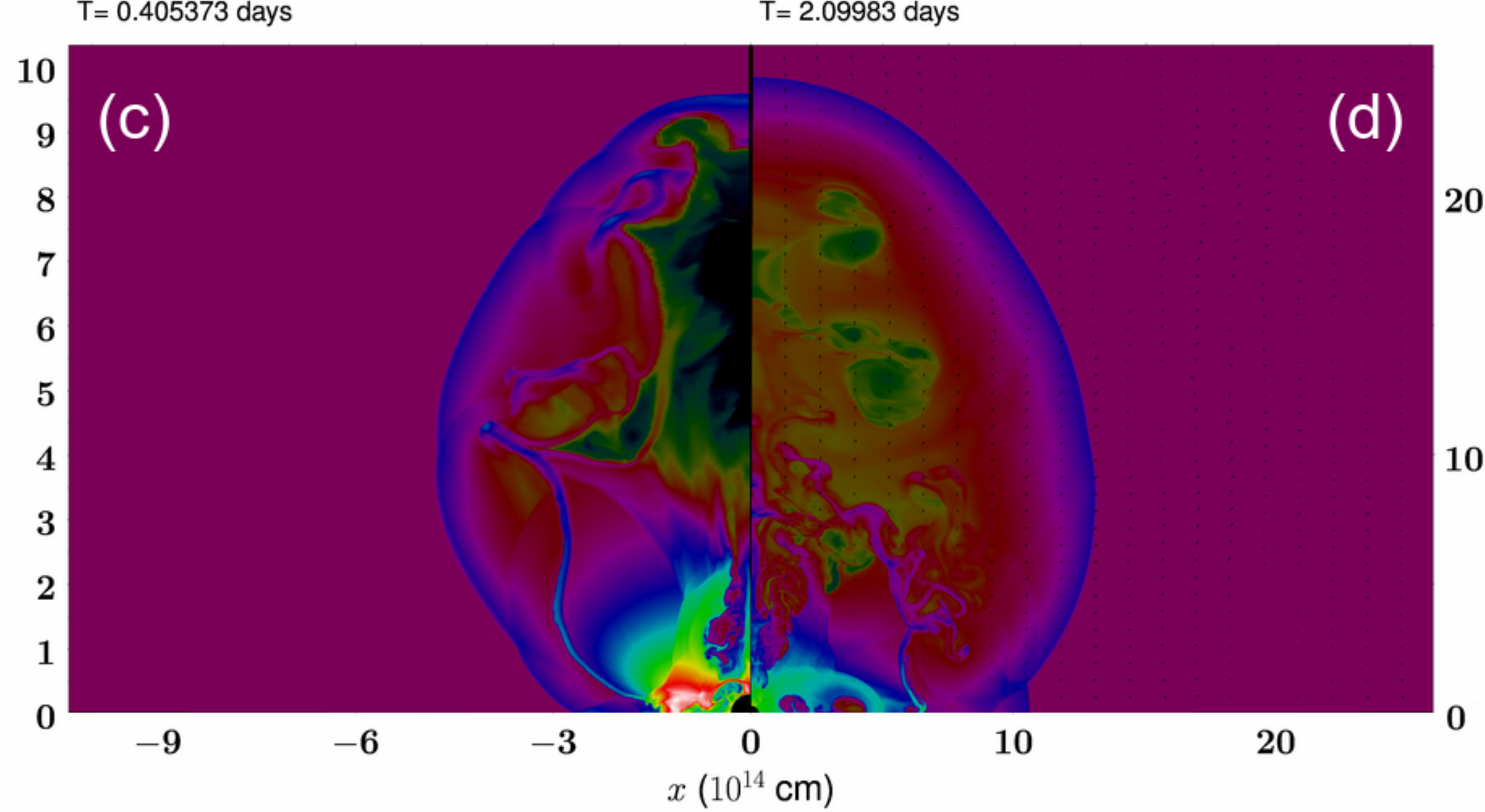}
\caption{Four snapshots of the rest-mass density evolution of the RM. The color scale is normalized to $\rho_{\rm ext}$. The time is displayed above each panel and refers to the laboratory frame time. 
}
\label{fig1}
\end{figure}

The jet dynamical evolution can de divided in three phases. In the first phase a jet is injected and accelerates before
it encounters the toroidal geometry of the high-density shell (Fig.\,\ref{fig1}a). In the second one, most of the jet interacts
with the CE shell while a part escapes through the funnel (Fig.\,\ref{fig1}b). The interaction produces a quick deceleration of
the jet, likely diminishing any standard afterglow signature. In the third phase, the jet material breaks out of the CE
shell, expanding sideways almost freely (Fig.\,\ref{fig1}c). The expansion is quasi-self-similar in the RM (Fig.\,\ref{fig1}d), but the
bubble dynamics depend on the external medium properties (see \cite{PaperI}).

\section{Temporal and spectral evolution}

\begin{figure}[h]
\includegraphics[height=.325\textwidth]{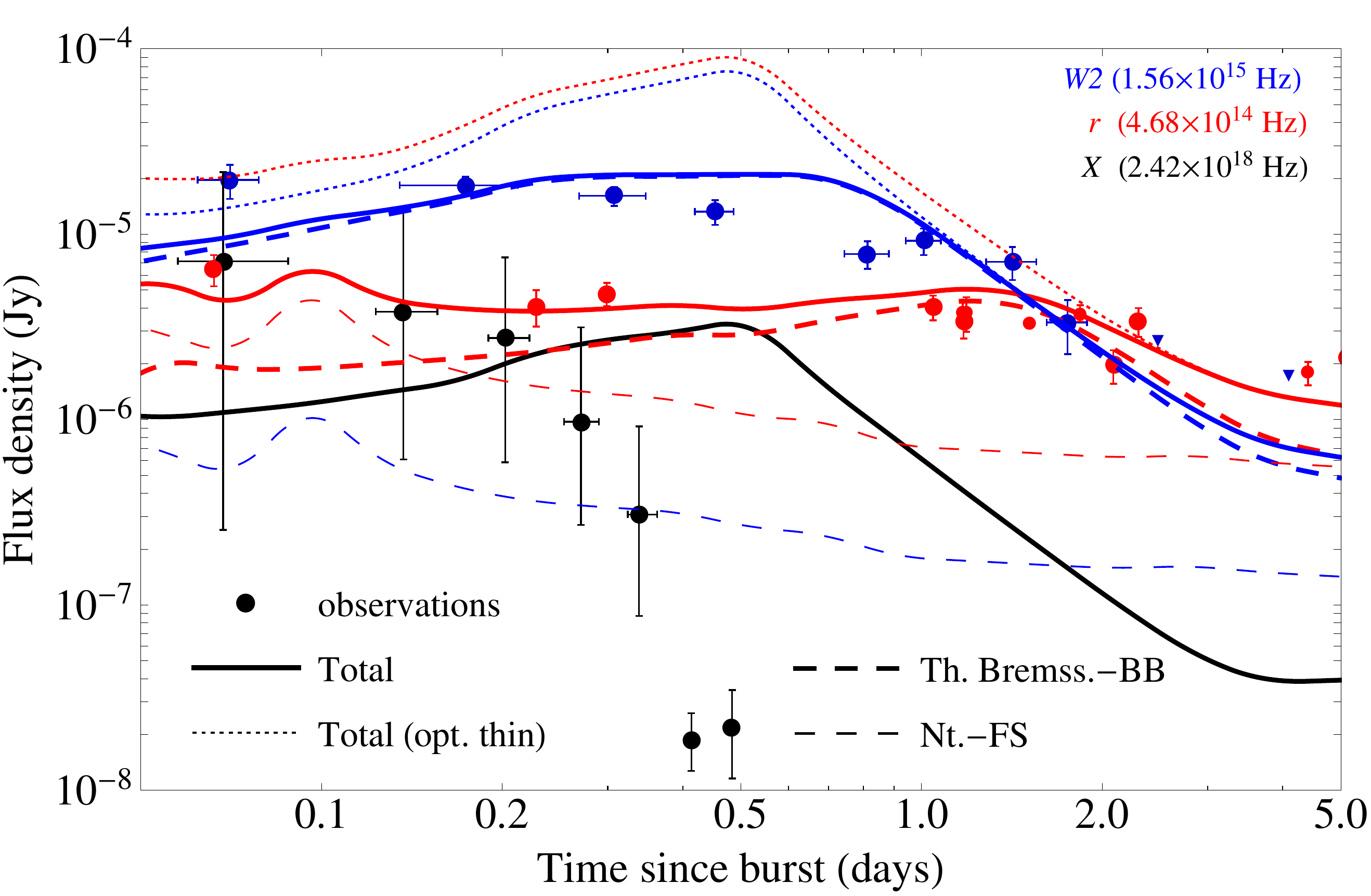}
\includegraphics[height=.325\textwidth]{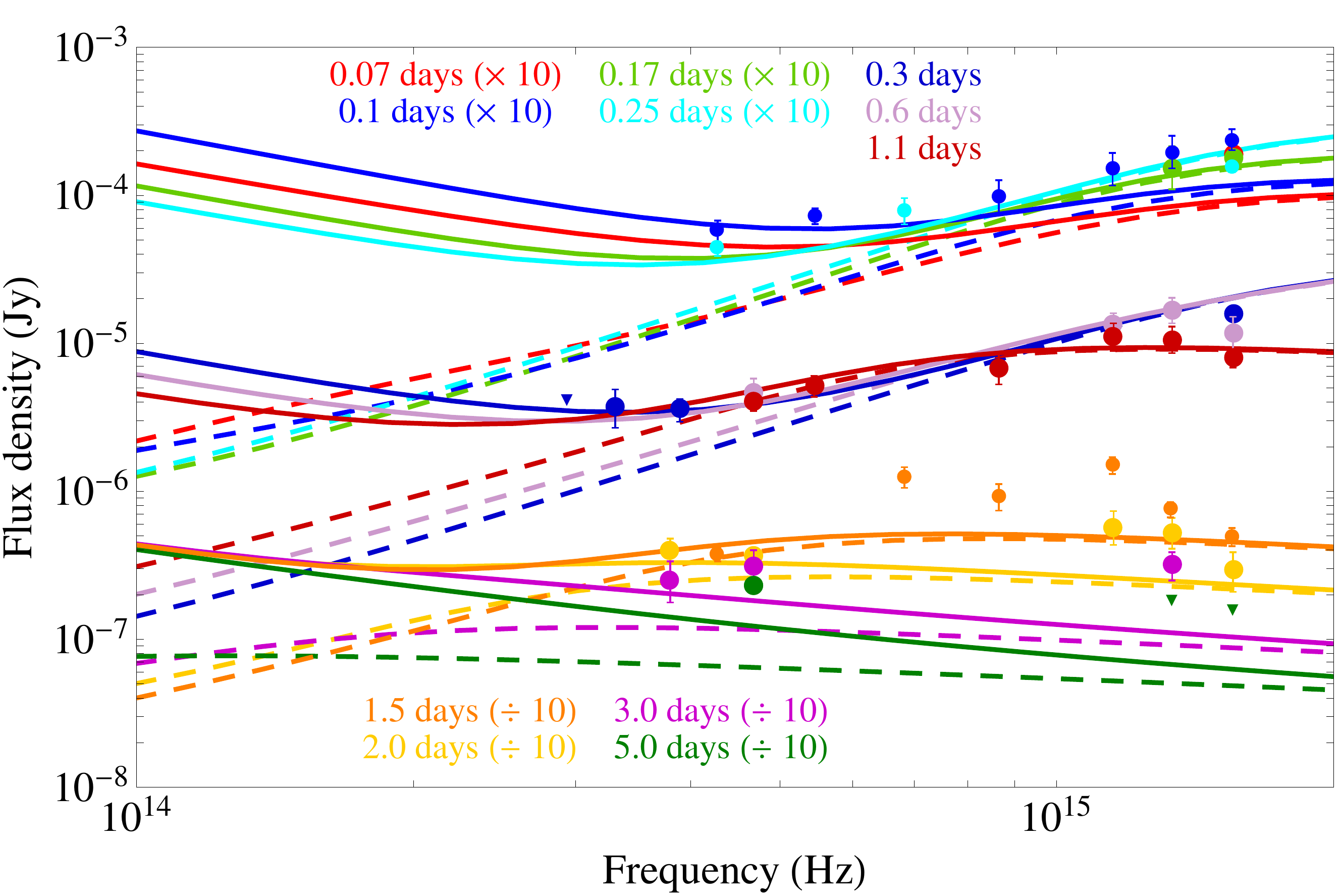}
\caption{Synthetic, optically thick light curves (left) and spectra (right) for the RM. We show the total emission (solid lines) and the individual contributions of the thermal (thick dashed lines) and the non-thermal radiation of the FS (thin dashed lines). We also plot the optically thin light curve for the total emission (dotted lines). Red, blue, and black colors in the left panel are used to display data in the $r$, $W2$, and X-ray bands, respectively. 
In the right panel, colors denote observations at different times (see legend). Note that for visualization convenience some of the data have been multiplied or divided by a factor of 10 (see the plot legends).
In both panels the observational data have been taken from \cite{T11} and references therein (large solid circles), and from \cite{L14} (small solid circles). Upper observational limits are represented as triangles.}
\label{fig2}
\end{figure}

We study the temporal evolution in two of the optical bands, the $W2$ and $r$ band, corresponding to frequencies of
$1.56\times10^{15}$ and $4.68\times10^{14}$ Hz, respectively. The synthetic emission of the RM reveals that this model
fits reasonably well the GRB 101225A observations up to $\sim5$ days (Fig. \ref{fig2}), as a combination of thermal radiation and
non-thermal radiation from the forward shock (FS). The former is the dominant contribution at higher frequencies and from
$\sim0.1-0.2$ days, and represents a $93\%$ and a $63\%$ of the total flux in the $W2$ and $r$ bands,
respectively, while the latter is dominant at low frequencies and during the early phases of the evolution, when the
shock is still relativistic. The emission of the reverse shock decreases rapidly since the jet sctructure is quickly
lost due to the strong interaction with the CE shell. Synchrotron emission from the CE-shell/jet shock is also
considered but it is negligible at the considered
frequencies. 
We also compute the X-ray emission and conclude that the thermal emission in this band ($2.42\times10^{18}$ Hz) is clearly
dominant, and that the estimated flux is only marginally consistent with observations until 0.3 days. We attribute this
discrepancy to the fact that the X-ray hotspot is too small to be resolved by our simulation, i.e. it is much larger in
our simulations than it should be according to observational estimates.

\section{Origin of the thermal radiation}

\begin{figure}[h]
\includegraphics[width=1.0\textwidth]{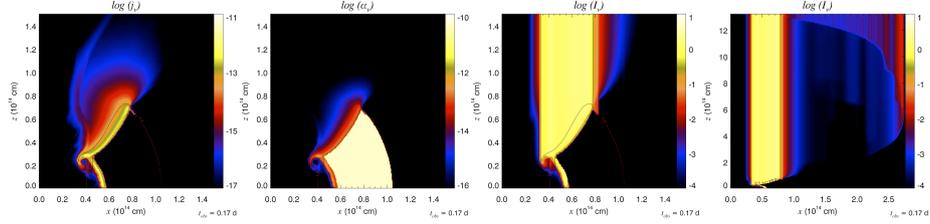}
\caption{Emission, $j_\nu$ (left),  absorption, $\alpha_\nu$, (left center) coefficients and evolution of the specific intensity, $I_\nu$ (right center) along the line of sight ($\theta_{\rm obs} = 0^\circ$) for free-free (thermal) bremsstrahlung process. The image zooms in the jet/CE-shell interaction region.
The total (thermal + non-thermal) specific intensity coming from the whole bubble along the line of sight is shown on the right panel. All the variables are in CGS units, calculated in the $W2$ band at an observational time $t_{\rm obs} = 0.17$ days.}
\label{fig3}
\end{figure}

\begin{figure}[h]
\includegraphics[width=1.0\textwidth]{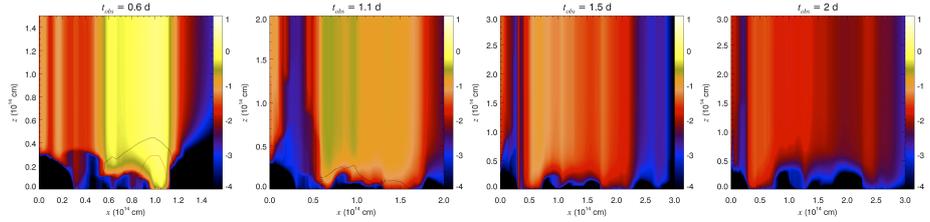}
\caption{Evolution of the specific intensity, $I_\nu$, in the $W2$ band (same as the right panels in Fig. \ref{fig3}). The image is focused on the jet/CE-shell interaction region. Note that the transition from optically thick to optically thin at $\sim1.5-2$\,days (right center and right panels) is due to the ablation of the CE-shell, which is absent after $\sim 2\,$days. The observational times are provided above of each of the panels. Optical depth contours of 1 (red line) and 0.1 (black line) are plotted.}
\label{fig4}
\end{figure}

We have shown that observations of the CB are mostly explained considering only thermal radiation. Thus, we have
specifically addressed where this emission is originated. We find that the thermal emission dominantly comes from
the interaction region between the jet and the CE shell (Fig. \ref{fig3}), showing that the presence of a very dense medium is
crucial for the model. Besides, the spectral inversion and reddening happening at $~1.5-2$ days can be related to the
time at which the massive CE shell is completely ablated by the jet (Fig. \ref{fig4}), while the system becomes optically thin
(Fig. \ref{fig2}, left). This feature is independent of the external medium properties, as we show in \cite{PaperI}.

\section{Conclusions}
We have seen that optical observations of the CB can be chiefly explained as thermal radiation from the CE-shell/jet interaction region.
In this way we have tested that the NS/He merger scenario is a plausible progenitor of BBD-GRBs since they produce the key element of the model: a high-density structure in the circumburst medium, i.e. the CE Shell. 
However, we do not rule out another possible progenitor scenarios \cite{L14,nakauchi13}. 

\section*{Acknowledgements}
We acknowledge the support from the European Research Council (grant CAMAP-259276), and the partial support of grants AYA2013-40979-P, CSD2007-00050 and PROMETEO-2009-103, and also the support of ACIF/2013/278 fellowship. 


\end{document}